\documentclass[twocolumn, prl]{revtex4-1}
\usepackage{graphicx}
\usepackage{amsmath}
\usepackage{longtable}
\usepackage{latexsym}
\usepackage{times}
\usepackage{graphics}
\usepackage{graphicx}
\usepackage{epsfig}
\usepackage{wrapfig}
\usepackage[T1]{fontenc}
\usepackage{float}
\usepackage{enumerate}
\usepackage{setspace,ifthen}
\usepackage{amsthm} 
\usepackage{amssymb}	
\usepackage[usenames,dvipsnames]{xcolor}
\usepackage{mathrsfs,amsmath}
\usepackage{mathrsfs}
\usepackage{units}
\usepackage{braket}

\date{\today}
\begin{document}
\title{Prediction and real-time compensation of qubit decoherence via machine learning}

\author{Sandeep Mavadia$^{*}$, Virginia Frey$^{*}$, Jarrah Sastrawan, Stephen Dona, Michael J. Biercuk$ ^{\dagger} $}
\affiliation{ARC Centre for Engineered Quantum Systems, School of Physics, The University of Sydney, NSW 2006 Australia}
\affiliation{National Measurement Institute, West Lindfield NSW 2070 Australia}
\email{\emph{These two authors contributed equally to this work} \newline $^{\dagger}$Contact: michael.biercuk@sydney.edu.au }

\begin{abstract}
The wide-ranging adoption of quantum technologies requires practical, high-performance advances in our ability to maintain quantum coherence while facing the challenge of state collapse under measurement. Here we use techniques from control theory and machine learning to predict the future evolution of a qubit's state; we deploy this information to suppress stochastic, semiclassical decoherence, even when access to measurements is limited.  First, we implement a time-division-multiplexed approach, interleaving measurement periods with periods of unsupervised but stabilised operation during which qubits are available, for e.g. quantum information experiments. Second, we employ predictive feedback during sequential but time delayed measurements to reduce the Dick effect as encountered in passive frequency standards.  Both experiments demonstrate significant improvements in qubit phase stability over ``traditional'' measurement-based feedback approaches by exploiting time domain correlations in the noise processes. This technique requires no additional hardware and is applicable to all two-level quantum systems where projective measurements are possible.

\end{abstract}

\maketitle

\section{Introduction}
The applications of quantum-enabled technologies are compelling and already demonstrating significant impacts, especially in the realm of sensing \cite{Gustavson1997,Chou2010,Shah2007,Aasi2013,Wasilewski2010} and metrology \cite{Leibfried2004}.  However, in nearly all applications the phenomenon of decoherence - effectively the randomisation of a quantum system's state by the environment - limits the viability of quantum technologies.  In the case of qubits, fundamental building blocks in many applications, the net result is that the useful lifetime of the qubit state is shortened, reducing their deployability for quantum information~\cite{NC2000}, quantum simulation~\cite{Lloyd:1996, PhysRevLett.82.5381, PhysRevA.65.042323, AspuruGuzik:2012, Bloch:2012, Blatt:2012}, or other applications.  Methodologies for stabilising qubits against decoherence represent a critical need in quantum technology. 

\begin{figure}[t]
	\includegraphics[scale=1]{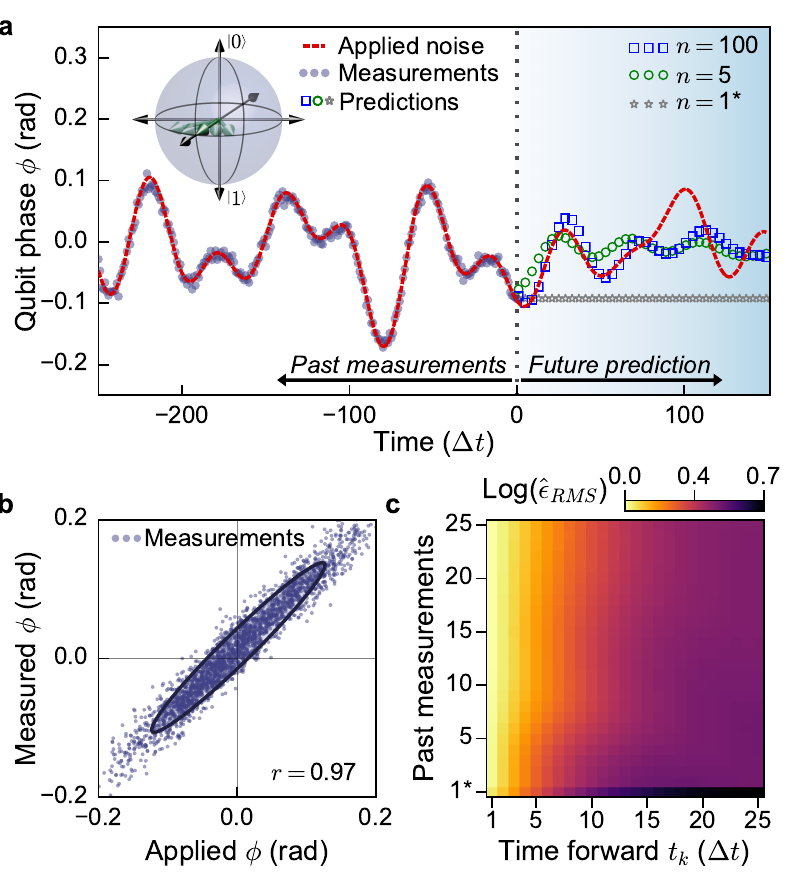}
	\caption{{Demonstration of the predictive algorithm on measurements of a qubit subject to engineered dephasing, resulting in randomisation of the qubit phase (inset).  \textbf{a}) The average phase evolution of the qubit during each measurement under the influence of an engineered noise trace, $\phi^{\textrm{A}}$, is probed via Ramsey spectroscopy and a projective measurement performed before the qubit state is reinitialised and the process repeated (see \emph{Supplementary Methods}). Time is represented in discrete increments of $\Delta t$, approximately corresponding to the measurement time. Values of $t_{k \leq 0}$ refer to past measurements used to make predictions and $t_{k > 0}$ refer to future predictions.
			Noise possesses a quasi-white power spectral density up to frequency cut-off $\omega_{\mathrm{c}}$, which we sample at $\omega_{\mathrm{s}} = 40 \omega_{\mathrm{c}}$.  Future qubit evolution is calculated offline based on these measurements.  Data labelled ``$n=1^*$'' correspond to traditional feedback (no prediction).   
			\mbox{\textbf{b}) 
				Correlation} between $\phi^{\mathrm{M}}\left(t_{k}\right)$ and $\phi^{\mathrm{A}}(t_{k})$ represented as a scatter plot for all measurements in this data set.  Ellipses are guides to the eye calculated to have major and minor axis determined by the eigenvectors of the data's covariance matrix.  The Pearson product-moment correlation coefficient, $r$, is calculated to quantify the quality of the measurements - here 97\%. }
		\mbox{\textbf{c}) Normalised} RMS errors $\hat{\epsilon}_{\mathrm{RMS}}$ between $\phi^{\mathrm{A}}(t_{k})$ and $\phi^{\mathrm{P}}\left(t_{k}\right)$ as a function of past measurements and discrete steps forward in time, averaged over all elements of the data set. Data are normalised to the lowest overall value in the field and are presented using a logarithmic scale to highlight differences over a broad dynamic range.  The first row corresponds to traditional feedback.}
	\label{Fig1}
\end{figure}

Control engineering~\cite{Stengel} techniques are emerging as a promising alternative to engineering passive robustness at the device level in realising stable quantum systems \cite{Tarn2003, Wiseman2005,James2009, BiercukJPB2011}. Beyond widely adopted open-loop control~\cite{Viola1998, BiercukJPB2011, Soare2014filtering}, a qubit subjected to stochastic evolution of its phase degree of freedom -- dephasing (inset Fig.~\ref{Fig1}a) -- can be stabilised by cyclically performing measurements on the qubit and then compensating for the measured phase evolution in a feedback loop~\cite{Bluhm_Feedback, Siddiqi2012, Shulman2014}. However, so far, feedback control ~\cite{Mabuchi2004, Kippenberg, Aspelmeyer, Sayrin2011, Gillett2010} has largely been limited by state-collapse under projective measurement, mandating access to weak measurements~\cite{Vijay2012} or ancilla states~\cite{Hirose2016}, or largely sacrificing useful quantum coherence in the controlled system~\cite{Shulman2014}.    



Our objective is to enhance the performance of incoherent feedback stabilisation (i.e. using only classical information) of a qubit experiencing dephasing while also relaxing the need for projective measurements.  Our approach is based on predictive control; a variety of techniques in filtering~\cite{Stengel, Landau2011, Mabuchi2009, Siddiqi2015} and machine learning~\cite{Murphy2012} allow the estimation of future state evolution based on past measurement outcomes of the system.  Here, we deploy a well established algorithm from machine learning to learn about a random dephasing process affecting a qubit, and then predict the impact of future dephasing based only on standard projective measurements.  We use this information to perform real-time stabilisation of the qubit state during periods in which the qubit is unsupervised but still subject to stochastic dephasing.  Our method exploits the presence of commonly encountered temporal correlations in the dephasing process~\cite{Clarke2004} in order to allow future prediction; no deterministic model of qubit state evolution is required.  To the best of our knowledge, despite its ubiquity in classical settings, predictive control has not been employed in the context of quantum-coherent technologies.

\section{Results}
\subsection{Supervised learning based on qubit-phase measurements}
In the language of machine learning, we consider the qubit's instantaneous phase which we would like to predict at a future discretised time, $t_{k}$, as labels, $\phi^{\mathrm{P}}\left(t_{k}\right)$, and an arbitrary number, $n$, of previous measurements, $\phi^{\mathrm{M}}_{i}$ (indexed by $i$ and obtained by any appropriate method), as their associated features.  We then calculate a linear combination of the features with optimised weighting coefficients, $\mathbf{w} = \{ w \}_{i,k}$, as a prediction of the label, $\phi^{\mathrm{P}}\left(t_{k}\right) = w_{0,k} + \sum^{n}_{i = 1} w_{i,k} \phi^{\mathrm{M}}_{i}$.  Based on measured features, the entries of $\mathbf{w}$ are optimised for each time step, $t_{k}$, reflecting the time-varying correlations in the dephasing process, captured through the power spectrum.
 
We demonstrate prediction of a qubit's state subject to stochastic dephasing by performing experiments using the ground-state hyperfine states, $\left|F=0,m_{\mathrm{F}}=0\right\rangle$ and $\left|F=1,m_{\mathrm{F}}=0\right\rangle$, in trapped $^{171}$Yb$^{+}$ ions as a qubit with transition frequency near 12.6\,GHz. A coherent superposition of the qubit states in the measurement basis induced by microwave control~\cite{Soare2014bath} evolves freely under the influence of an engineered dephasing interaction larger than any intrinsic noise in our experimental system (see \emph{Supplementary Methods}).  In general we work in a regime where the noise evolves slowly during a single measurement period $T_{\mathrm{M}}$, but we allow the rate at which measurements of qubit phase evolution are taken - the sampling frequency $\omega_{\mathrm{s}}$ - to vary relative to the highest frequency in the noise power spectrum, $\omega_{\mathrm{c}}$ (c.f. Fig.~\ref{Fig3}f). The dephasing noise processes presented here are all derived from a flat-top frequency power spectrum with characteristic cut-off at $\omega_{\mathrm{c}}$. More complex spectra are discussed in the \emph{Supplementary Discussion} and demonstrate similar performance.

An important aspect of our approach is that measurements providing data serving as features may be performed through any suitable protocol.  For instance, performing a series of $p$ projective measurements on a single qubit in order to obtain ensemble-averaged information simply sets the scale of the measurement period, $T_{\mathrm{M}}\to pT_{\mathrm{M}}^{(1)}$, with $T_{\mathrm{M}}^{(1)}$ the duration of a single experiment.  Here, we employ a projective measurement that captures statistical information through a spatial ensemble.  The impact of such differences is explicitly captured in the sampling frequency of the measurement process.

\subsection{Forward prediction of stochastic qubit phase evolution}
We begin by accumulating a series of projective measurements of the qubit's phase under engineered dephasing.  These serve as training data for the algorithm to optimise the coefficients in $\mathbf{w}$.  We then perform another series of measurements (shown, Fig.~\ref{Fig1}a) under application of a different noise process possessing similar statistical characteristics as used in acquiring the training data. This approach ensures that our estimates of prediction accuracy are conservative and exhibit reasonable model robustness and generality.  Performing the learning algorithm on a single data set can enhance performance of the prediction algorithm but introduces extreme sensitivity to the input model, ultimately reducing prediction efficacy in the presence of variations in the detailed form of the noise.   

An example engineered noise trace in time with overlaid measurement outcomes, $\phi^{\mathrm{M}}$, is depicted in Fig.~\ref{Fig1}a, with 97\% correlation between $\phi^{\mathrm{M}}$ and the applied phase $\phi^{\mathrm{A}}$ (Fig.~\ref{Fig1}b).  Beyond time $t_{0}$ we predict future labels of qubit phase evolution $\phi^{\mathrm{P}}\left(t_{k}\right)$, up to step $t_{150}$ using a variable number, $n$, of past measurements and the trained coefficients in $\mathbf{w}$.  Calculated predictions approximate $\phi^{\mathrm{A}}$ well, reproducing key features including inflection points, maxima, and minima as a function of $t_{k}$.  Our knowledge of the noise is used exclusively for quantitative evaluation of prediction efficacy - it does not enter into the machine learning algorithm in any form. 

Prediction accuracy increases with $n$, as the algorithm learns more about the temporal correlations in $\phi^{\mathrm{A}}$. For values of $k\gtrsim n$, corresponding to prediction times exceeding the range over which the algorithm possesses knowledge about the noise features, the prediction quality diminishes. In addition, over very large values of $t_k$ the prediction tends towards the mean of the noise. Comparing predictive estimation to a ``traditional feedback'' model, in which future estimates are based simply on the last measured value $\phi^{\mathrm{M}}(t_0)$, the algorithm shows a distinct advantage as it allows for temporal evolution of the noise in the future.  

The quantitative benefits of predictive estimation relative to traditional feedback, and the large $t_{k}$ behavior of the predictive algorithm are succinctly captured in the root-mean-square (RMS) prediction error averaged over the entire dataset, $\epsilon_{\mathrm{RMS}}$, and calculated as a function of $t_{k}$ and $n$ (Fig. \ref{Fig1}c).  This demonstrates that even over a large ensemble of predictions the algorithm's advantages remain robust.  We now move on to provide examples of real-time qubit stabilisation in which the incorporation of future state prediction shows significant advantages over existing techniques. 
 


\begin{figure}[ht]
\includegraphics[scale=1]{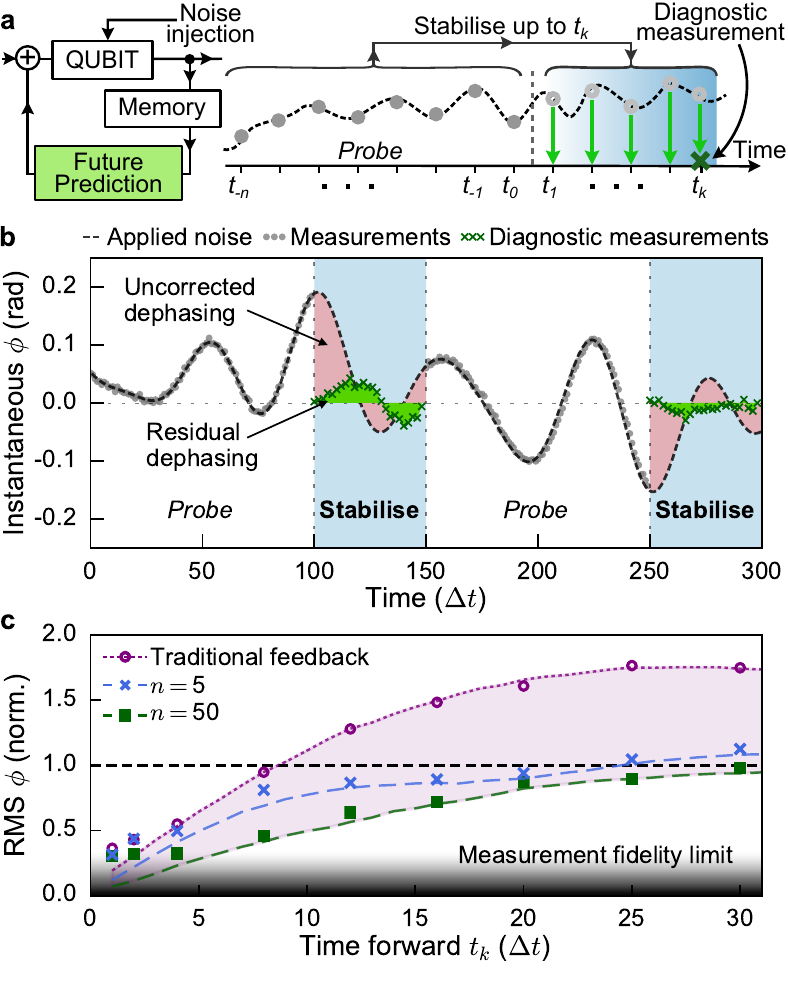}
\caption{Experimental time-division multiplexing for qubit stabilisation against dephasing. \mbox{\textbf{a}) Schematics} showing the key aspects of the implementation. Noise is continuously injected into the system. Measurements are taken up to \mbox{$t_{0}$} and processed in real-time to predict future evolution of the qubit phase many time steps ahead, $\phi^{\mathrm{P}}(t_{k})$.  From $t_{0}$ measurement-free compensation based on $\phi^{\mathrm{P}}(t_{k})$ is applied during each discrete time step (light green arrows) up to $t_k$ when a diagnostic measurement is performed to verify the accuracy of the prediction/correction process. Full details appear in the \emph{Supplementary Methods}.
\mbox{\textbf{b}) Probe} and stabilisation cycles of a time division multiplexed (TDM) measurement using \mbox{$n=100$} past measurements and prediction/correction up to $k=50$ time steps ahead. Green shading indicates reduced residual phase errors.  
\mbox{\textbf{c}) RMS} results from TDM measurements for different $t_{k}$ and $n$ compared against traditional feedback and averaged over 50 unique stabilisation periods. Data are normalised to the RMS of $\phi^{\mathrm{A}}$ indicated by the horizontal dashed line. The other dotted/dashed lines are simulations.  Markers represent the averaged results of diagnostic measurements. For these data $\omega_{\mathrm{s}}=40\omega_{\mathrm{c}}$.}
\label{Fig2}
\end{figure}



\subsection{Time-division-multiplexed decoherence suppression}
 As described above, a reliance on feedback involving frequent projective measurements renders a qubit effectively useless for quantum information or other applications, but omission of stabilisation techniques in the presence of dephasing noise may lead to phase errors and eventually to total decoherence.  To mitigate the effect of dephasing, we tailor an approach in which we temporally multiplex the necessary measurement and actuation operations in distinct probe and stabilisation periods respectively (Fig.~\ref{Fig2}a, b).   During the probe period, a fixed number of measurements are taken and processed in real time.  From these measurement outcomes the algorithm produces a prediction of the future time-dependent evolution of the noise during the subsequent stabilisation period up to some $t_{k}$;  the qubit is dedicated exclusively to measurement of the dephasing process in the probe period.  During the stabilisation period, corrections are applied during each discrete time step to compensate the predicted stochastic phase evolution, but no measurements are conducted; this permits periods of unsupervised evolution during which the qubit is useful and stabilised against dephasing.  

As an example we set the objective of maintaining zero net qubit phase accumulation (in the rotating frame) during each timestep of the stabilisation period such that arbitrary high-fidelity operations may be conducted on the qubit; here we apply only the identity.  Diagnostic measurements are performed after a variable number of corrections in order to demonstrate the efficacy of this approach but would not ordinarily be required. Two representative probe/stabilisation cycles are displayed in Fig.~\ref{Fig2}b showing a reduction in integrated phase error of about 70\% after a stabilisation delay of $t_{50}$ during the first cycle and a reduction of about 85\%  during the second. These improvements are partially limited by measurement fidelity, as illustrated in the ensemble-averaged data (Fig.~\ref{Fig2}c).  Predictive compensation in all tested regimes is superior to corrections based only on traditional feedback down to measurement fidelity limits.   Compared against numerical simulations we see that for small $t_{k}$ the algorithm can provide large relative gains. 


\subsection{Predictive estimation inside a periodic feedback loop}
In a second application we employ real-time predictive control in a metrological context.  Qubits realised in atoms are frequently used as stable references against which local oscillators (LOs) may be disciplined~\cite{audoin2001}.  However, stochastic evolution of the LO frequency between interrogations leads to imperfect corrections in the feedback loop.  This scenario is commonly encountered when classical processing, actuation, and system reinitialisation introduce dead time, producing an effective lag in the feedback loop which degrades the long-term stability of the locked oscillator~\cite{greenhall1998}.  The impact of rapid fluctuations in the LO frequency relative to dead time is generally referred to as the Dick effect~\cite{dick1987}, and represents a significant limiting phenomenon in passive frequency standards using atomic references.  The correspondence between LO-induced instabilities in frequency references and dephasing in qubits~\cite{Ball2016} thus invites the application of predictive control in a setting where periodic interrogation and projective measurement are native to the feedback loops used in precision frequency metrology.

\begin{figure}[t!]
\includegraphics[scale=1]{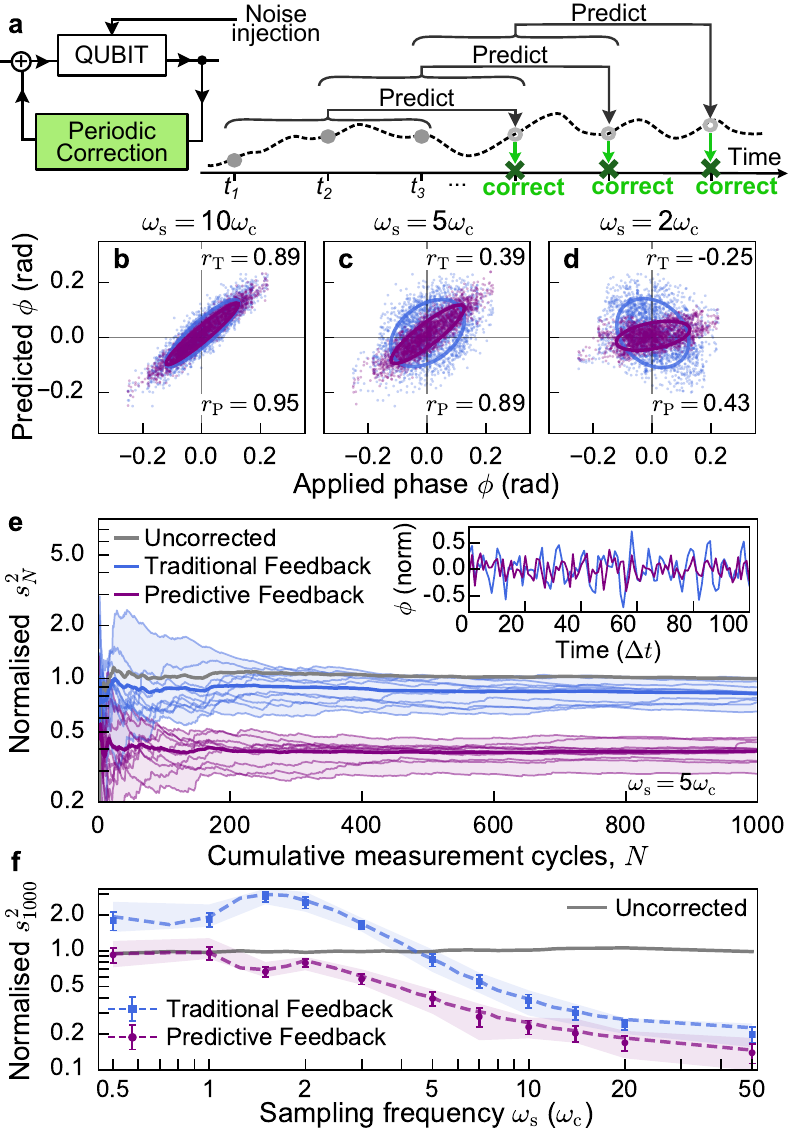}
\caption{Experimental comparison of long-term stabilisation using traditional and predictive feedback. 
\mbox{\textbf{a} Schematics} showing the key aspects of our cyclic feedback implementation using overlapping measurements.
\mbox{\textbf{b-d} Demonstration} of feedback accuracy for different sampling frequencies $\omega_{\mathrm{s}}$ quantified in units of $\omega_{\mathrm{c}}$, presented through correlation plots (c.f. Fig. \ref{Fig1}c) for traditional feedback (blue) and prediction (magenta).  Data presented are derived from Fig.~\ref{Fig1}a.
\mbox{\textbf{e} Measured} sample variance for various protocols as a function of the number of cycles.  Data are normalised to the sample variance of the uncorrected (free-running) signal at 1000 samples.  Each line represents data taken for one particular noise realisation and thick lines represent the ensemble average.  The inset shows an example suppression of variance over measurement outcomes using predictive against traditional feedback (normalised to the noise amplitude).
\mbox{\textbf{f}  Sample} variance at $N=1000$ as a function of sampling frequency $\omega_{\mathrm{s}}$ in units of $\omega_{\mathrm{c}}$, normalised to the sample variance of the uncorrected signal. The measurement time is fixed and $\omega_{\mathrm{s}}$ varied through introduction of dead-time between measurements.  Dotted lines display simulations and markers the measurement results averaged over ten noise realisations. Error bars represent the standard deviation of the mean and the shaded areas show the maximum spread of outcomes. For fixed noise parameters varying $\omega_{\mathrm{s}}$ serves as a proxy for changing the ratio of $T^{-1}_{\mathrm{M}}/\omega_{\mathrm{c}}$ (see \emph{Supplementary Discussion}).  Simulations and measurements in all panels use $n=20$.}
\label{Fig3}
\end{figure}

The usefulness of predictive estimation in improving correction accuracy inside a feedback loop is demonstrated in Fig.~\ref{Fig3}b-d, where we plot the predicted phase $\phi^{\mathrm{P}}(t_{k})$ (based on two different techniques) against the applied phase error $\phi^{\mathrm{A}}(t_{k})$.  A prediction with unity correlation to the applied noise would form a diagonal line along $\phi^{\mathrm{P}} = \phi^{\mathrm{A}}$ (similar to Fig.~\ref{Fig1}b), while imperfect predictions - hence imperfect corrections - result in a dispersion of points around this line in an ellipse. 

We vary the sampling frequencies $\omega_s$ as a proxy for introducing a variable dead time in the feedback loop (see \emph{Supplementary Discussion}). In a regime where the LO-induced dephasing process evolves slowly, quantified as \mbox{$\omega_{\mathrm{s}} \gg \omega_{\mathrm{c}}$}, both $\phi^{\mathrm{M}}\left(t_{0}\right)$ and the predicted phase $\phi^{\mathrm{P}}(t_{k})$ show positive correlation to $\phi^{\mathrm{A}}(t_{k})$ (Fig.~\ref{Fig3}b).  As we decrease $\omega_{\mathrm{s}}$, noise evolution during the dead time leads to diminishing correlation between the prediction and actual noise, causing the ellipses to rotate and broaden - a manifestation of the Dick effect. 

Predictive estimates are compared to the traditional feedback model described above.  For $\omega_{\mathrm{s}}$ approaching the Nyquist limit we observe that the traditional prediction can become anticorrelated with the rapidly evolving applied noise (blue ellipse, Fig.~\ref{Fig3}d), which in real-world applications would lead to an unstable system under feedback.  By contrast, using optimised predictions, the decrease in correlation is much slower and the machine learning algorithm prevents the prediction from ever becoming anticorrelated with the applied dephasing noise. In circumstances tested we always find the optimal prediction correlation $r_{\mathrm{P}}>r_{\mathrm{T}}$ for traditional feedback.  Corrections used to discipline the qubit or LO based on predictive estimation can therefore possess enhanced average accuracy relative to traditional feedback.  

We now implement real-time evaluation of $\phi^{\mathrm{P}}(t_{k})$ inside a feedback loop, demonstrating the ability to improve the individual corrections and ultimately achieve improved long-term stability of the locked qubit.  In our experiment we set $n=20$, calculate $\phi^{\mathrm{P}}(t_{k})$ on the fly, and cyclically correct based on these predictions (Fig.~\ref{Fig3}a), again comparing against traditional feedback.  The long-term stability achieved under both methods is calculated via the sample variance~\cite{rutman1978} over a variable number of feedback cycles (Fig.~\ref{Fig3}e).  

Over the range of dead times explored experimentally, the use of optimised predictive feedback, in which future estimates are updated as new measurements are acquired in real time, yields net enhancements over the free-running LO (Fig.~\ref{Fig3}e, f).  This includes regimes near the Nyquist limit where rapid evolution of the noise can result in feedback-induced instability as in Fig.~\ref{Fig3}d.  Over most of this range and for the noise parameters we have employed, performance gains over traditional feedback are approximately $2\times$ using optimised predictive feedback - a metrologically significant improvement using only enhanced software in the stabilisation.  Similar performance enhancements have been observed for a wide range of noise spectra and parameters (see \emph{Supplementary Discussion}).

\subsection{Predictive estimation applied to intrinsic system noise}
\begin{figure}[t!]
\includegraphics[scale=1]{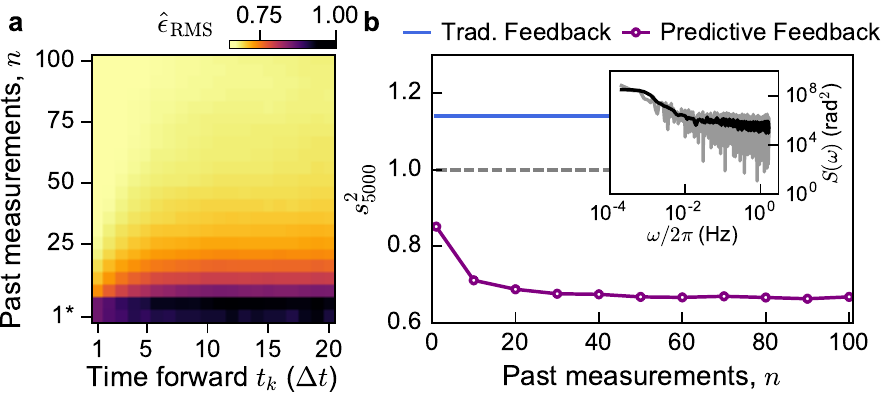}
\caption{Application of predictive qubit state estimation to intrinsic system noise. \textbf{a}) RMS errors between predictions, $\phi^{\mathrm{P}}$ and actual values $\phi^{\mathrm{P}}$ for various numbers of past measurements and discrete steps forward in time, averaged over the whole set of validation data. The RMS values are normalised to the RMS deviation of the uncorrected data from zero. The bottom row ($1^*$) corresponds to traditional feedback. \textbf{b}) Sample variance of the corrected measurements averaged over 5000 cycles, as a function of past measurements used for prediction, normalised to the sample variance of the uncorrected system. The expected sample variance obtained by performing traditional feedback is added for comparison. Data are split into two subsets, where the first $\unit[70]{\%}$ serve for training purposes and the remaining $\unit[30]{\%}$  are used for validation. \textbf{Inset}) Power spectrum of a series of projective measurements on the free-running qubit-LO system. The data is overlaid with a smoothed version to visualise the general trend. The maximum frequency in the spectrum corresponds to our sampling frequency and is about $\unit[1.7]{Hz}$.}
\label{Fig4}
\end{figure}

Finally, with quantitative evaluation of these techniques in hand using engineered noise, we move on to a study of the intrinsic dephasing noise in our system, which arises due to a combination of LO phase noise and magnetic field fluctuations. 
We perform thousands of sequential projective measurements on the free-running qubit-LO system and process predictions offline. The spectrum of measured fluctuations combines a $1/f^2$ type low-frequency tail with an approximately white plateau, resulting in significant spectral weight near the measurement cycle time.   We perform an analysis similar to that presented in Fig. \ref{Fig1}, with prediction accuracy quantified using the RMS error between predictions and the future measurement outcomes as a function of $t_{k}$ (Fig. \ref{Fig4}a). 

Our machine learning algorithm enhances the prediction of future qubit evolution by approximately $30\%$ relative to the RMS error of the uncorrected measurements. We achieve similar performance gains relative to both traditional feedback and the free-running system in calculated sample variance over thousands of correction cycles based on predicted qubit phase, Fig.~\ref{Fig4}b. In this case the rapid evolution of the noise causes traditional feedback to produce a larger sample variance than free evolution - a situation similar to that experienced in Fig.~\ref{Fig3}d. The calculated performance enhancements of our method on the intrinsic system noise are significant and show that our algorithm possesses the capability to improve the stability against the noise background in our system. 


\section{Discussion}
In this work we have demonstrated the ability to deploy machine learning techniques to predict and pre-emptively compensate for stochastic qubit dephasing.  By exploiting temporal correlations in noise processes, we are able to suppress dephasing during periods when probing the qubit state is not possible, even though we have no deterministic model of the qubit's evolution.  Implementing this approach requires neither additional quantum resources nor extra experimental hardware. Instead we rely on software-based machine learning techniques, which extract optimal performance from information that would have already been collected during common experimental implementations.  It has been shown numerically that it is possible to implement an analytical solution to maximally exploit noise correlations captured through the noise power spectrum~\cite{sastrawan2016}. However in our experimental demonstration the ease of implementation lends itself to use for large values of $k$ and $n$ where prediction is extended far into the future and the computational requirement of large matrix inversions make analytic techniques impractical.  In addition, deviations from the idealisation of noise characteristics represented by use of a simple power spectral density, as well as correlations appearing in the measurement process, are easily captured by the machine learning algorithm but invisible to such analytic approaches.


The capability to suppress errors in quantum systems undergoing stochastic evolution has direct implications for the metrology and quantum information communities.  In particular the ability to suppress the magnitude of residual dephasing errors makes this technique an attractive complement to open-loop dynamic error suppression for quantum information.  Any reduction in the strength of the effective noise experienced by the qubit exponentially improves the fidelity of an operation implemented using dynamic error suppression~\cite{Soare2014filtering}.  Even in the limit of quasi-static noise, reducing the magnitude of the dephasing error experienced during a dynamically protected operation will improve the ultimate fidelity achievable in a nontrivial quantum logic operation~\cite{Kabytayev2014}.  The complementarity between open- and closed-loop stabilisation is a common theme in control engineering and translates well to the current setting. Future experiments will involve an expansion to a greater variety of machine-learning algorithms for system characterisation and stabilisation, and treatment of more complex control scenarios with non-commuting noise terms in the qubit Hamiltonian, non-linearities in the control, and use of various measurement bases.

\bibliography{statepredict}
\bibliographystyle{ieeetr}

\section{End Notes}

\subsection{Author Contributions}
S.M. and V.F developed experimental hardware and the experimental control system, performed numerical simulations, obtained the presented data, and carried out the data analysis with guidance from M.J.B.  M.J.B. conceived the experiment and led development of the experimental system. J.S. and S.D obtained preliminary data. S.M., V.F. and M.J.B. wrote the manuscript.

\subsection{Data Availability}
Data published in this article and the computer code used for simulation is available from the authors.

\subsection{Competing Interests}
The authors declare no competing financial interests.

\subsection{Acknowledgements}
The authors acknowledge D. Hayes, M.C. Jarratt, and A. Soare, for contributions to the experimental system and C. Edmunds, C. Ferrie, and C. Hempel for useful discussions.  Work partially supported by the ARC Centre of Excellence for Engineered Quantum Systems CE110001013, ARC Discovery Project DP130103823, the Intelligence Advanced Research Projects Activity (IARPA) through the ARO, the US Army Research Office under Contract W911NF-12-R-0012, and a private grant from H. \& A. Harley.

\end{document}